\documentclass[pre,twocolumn,showpacs]{revtex4}
\usepackage{epsfig}

\begin{document}

\title{Enhanced rise of rogue waves in slant wave groups}
\author{V. P. Ruban}
\email{ruban@itp.ac.ru}
\affiliation{Landau Institute for Theoretical Physics,
2 Kosygin Street, 119334 Moscow, Russia} 

\date{\today}

\begin{abstract}
Numerical simulations of fully nonlinear equations of motion for long-crested
waves at deep water demonstrate that in elongate wave groups the formation of 
extreme waves occurs most intensively if in an initial state the wave fronts 
are oriented obliquely to the direction of the group. An ``optimal'' angle,
resulting in the highest rogue waves, depends on initial wave amplitude and 
group width, and it is about 18-28 degrees in a practically important range 
of parameters.
\end{abstract}
%47.35.Bb Gravity waves
%92.10.-c Physical oceanography
%02.60.Cb Numerical simulation; solution of equations
\maketitle

The phenomenon of rogue waves (also known as freak, killer, giant, or extreme waves) 
at the ocean surface has attracted much attention in recent years
(an extensive discussion can be found in  
\cite{KhPel2003,EJMBF2006,DKM2008,EPJST185,NHESS2010}, 
and particular aspects of the rogue wave formation are considered, for instance, in 
\cite{OOSB01,OOFS03,J03,P,LP06,FGD07,S-J05,GT07,On09,OOS06,SKEMS06,TB02011,ES2010,CHA2011}). 
A probable scenario suggests that linear mechanisms, such as interaction 
of surface waves with a nonuniform current, 
cause preliminary amplification of wave amplitude, making the most wide 
and tall wave groups unstable with respect to the so called modulational instability
\cite{Benjamin-Feir,Zakharov68,McLean_et_al_1981}. A rogue wave is thus 
the final stage of the development of that instability, as it has been confirmed
by direct numerical simulations of exact equations of motion for potential
two-dimensional (2D) flows of a perfect fluid with a free surface
\cite{ZDV2002,DZ2005,ZDP2006,DZ2008}. However, many three-dimensional (3D) aspects
of the problem still remain far from being clear. Partly the difficulty 
is explained by the absence of compact and explicit exact equations of motion
for a 2D free surface in the 3D space, which fact results in rather slow and cumbersome 
implementations of the existing numerical methods based on the Euler equations. 
Therefore some approximate analytical and numerical models were suggested 
to study 3D dynamics of oceanic waves. In particular, for weakly nonlinear 
regimes the nonlinear Schroedinger equation (NLSE) \cite{Zakharov68}
and its generalizations \cite{OOS06,SKEMS06,Dysthe1979,TD1996,TKDV2000} are widely used, 
which are simplifications of the Zakharov equation taking into account renormalized 
$2\to 2$ wave processes, which equation is the most general among weakly nonlinear 
models (see \cite{Zakharov68,Z1999}, and references therein). 
But weakly nonlinear equations are definitely not appropriate to describe rogue waves at 
their final stage of evolution. That is why another, fully nonlinear approximate model has 
been developed by the present author, based on a different small geometrical parameter,
which is a smallness of deviation from a planar flow \cite{R2005PRE}. In other words,
a narrow angular distribution of the wave spectrum in the horizontal Fourier plane is assumed.
The model is good for arbitrary steep long-crested water waves propagating closely 
to $x$ direction in the horizontal $(x,q)$-plane ($y$ is the vertical coordinate).
The corresponding numerical method is reasonably fast \cite{RD2005PRE}, and it
was applied to study breathing rogue waves in a random wave field  \cite{R2006PRE},
nonlinear stage of the modulational instability with specific zigzag coherent
structures producing freak waves through mutual interactions \cite{R2007PRL}, and
the two kinds of rogue waves in weakly-crossing sea states \cite{R2009PRE}.
Additional numerical examples can be found in the recent paper \cite{R2010EPJST}.

In the present work the author continues investigation of 3D effects 
in the dynamics of extreme water waves.
Let us consider sea states in situations when a typical wave has a length $\lambda$ 
and an amplitude $A$, so a typical wave steepness is $s=2\pi A/\lambda$. 
As a general rule, a width of the wave group $N\lambda$ should be sufficiently large 
for the rogue wave formation to take place,
\begin{equation}
\label{BFI}
sN\gtrsim I_c,
\end{equation}
with some constant $I_c\sim 1$ (precise value of $I_c$ is not important, since we speak 
now about typical values; besides that, there exist several slightly
different definitions for rogue waves; see discussion in \cite{EPJST185}). 
The product $sN$ is the so called Benjamin-Feir Index (BFI) \cite{J03}. 
With small BFI, dispersive effects in the wave dynamics are dominant, 
while for large BFI the nonlinearity becomes essential.
The above condition comes from simplified consideration of the phenomenon within 
one-dimensional (1D) NLSE describing a complex wave envelope $B(x,t)$ 
in the case of a planar flow (see \cite{Zakharov68}), 
\begin{equation}\label{NLSE_1D}
\frac{i}{\omega}\frac{\partial B}{\partial t}
+\frac{i}{2k}\frac{\partial B}{\partial x}=
\frac{1}{8k^2}\frac{\partial^2 B}{\partial x^2}+\frac{k^2}{2}|B|^2B,
\end{equation}
where $k=2\pi/\lambda$ is the wave number, $\omega=\sqrt{gk}$ is the wave frequency,
and $g$ is the gravitational acceleration. Condition (\ref{BFI}) means 
that the wave group contains in some sense a soliton of the 1D NLSE. By the way, 
the parameter $N$ is the number of individual waves in the 1D group.
One of the most intriguing questions in the theory of 3D rogue waves is about their
appearance in non-coherent random sea states, when typical wave groups are not very 
tall and/or wide, so  $sN\lesssim I_c$. 
However, in the cases when envelopes $B(x,q,t)$ of wave groups have
a length much longer than the width (elongate wave groups, as in weakly-crossing sea
states \cite{R2009PRE}; see also \cite{SB2010}), 
then an additional important parameter comes into play, 
namely an angle $\theta$, at which wave fronts are oriented relatively to the ``long'' 
direction. 
The purpose of the present work is to demonstrate via numerical simulations 
that in this regime the most high freak waves arise for some optimal angle $\tilde\theta$,
which depends on the parameters $N$ and $s$. This essentially three-dimensional effect 
is reported for the first time, and it is very distinct in practically important ranges 
$0.10\lesssim s\lesssim 0.14$ and $5\lesssim N \lesssim 10$. 
In this parametric region we find $\tilde\theta=18\cdots 28^o$, 
and  arising rogue waves have the height $Y_{\rm max}\approx 0.06\lambda$ 
at which the process of wave breaking  begins to occur, while the ratio 
$Y_{\rm max}/A=(2.5\cdots 3)$. At the same time, $Y_{\rm min}\approx -0.04\lambda$,
due to the crest-trough asymmetry of gravity water waves.  

To deal with a minimal set of parameters in our study, we consider idealized wave 
groups which are infinitely long in one direction. So we have stripes along $x_2$ 
axis in a turned $(x_1,x_2)$ coordinate system which is oriented at some angle 
$\gamma$ to the $(x,q)$-system, with $\gamma$ being slightly less than $\theta$. 
The components of the corresponding wave vector in $(x_1,x_2)$ coordinate system are
$(k\cos\theta,-k\sin\theta)$. Thus, initial wave fronts are oriented not exactly 
parallel to $q$-axis, but at a small angle (about several degrees) clockwise,
while the stripe itself is oriented at the angle $\gamma$ anticlockwise.
This is made because crests of arising extreme waves are always oriented more gently 
to the stripe direction comparatively to the crests in the initial state 
(see Fig.1, and also \cite{R2007PRL,R2010EPJST}), so the choice $\gamma<\theta$ 
results in a more close orientation of rogue wave crests to $q$-axis,  
as it is required for applicability of the employed approximate quasi-2D model
\cite{R2005PRE,RD2005PRE}.

In the initial state, the complex envelope of the first wave harmonics is put 
purely real and given by a simple expression,
\begin{equation}
B(x_1,0)\approx\frac{s}{k}\exp\left(\frac{-x_1^2}{2w^2\lambda^2}\right).
\end{equation}
Thus, we can identify the parameter $N$ as follows: $N\approx 4w$. 
We also add a low-level random-phase perturbation into initial wave spectrum, 
similarly to \cite{R2006PRE,R2007PRL}, to be sure that our results are robust
with respect to a noise in initial conditions. 

\begin{figure}
\begin{center}
   \epsfig{file=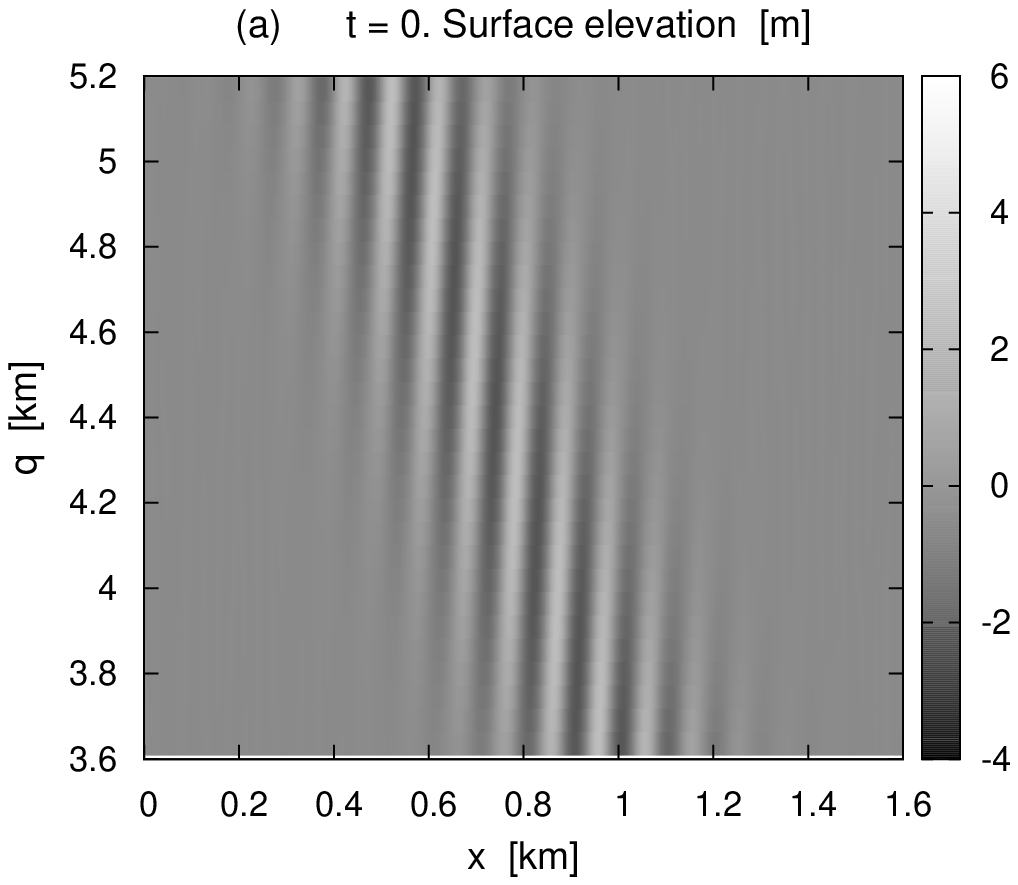,width=91mm}\\
   \epsfig{file=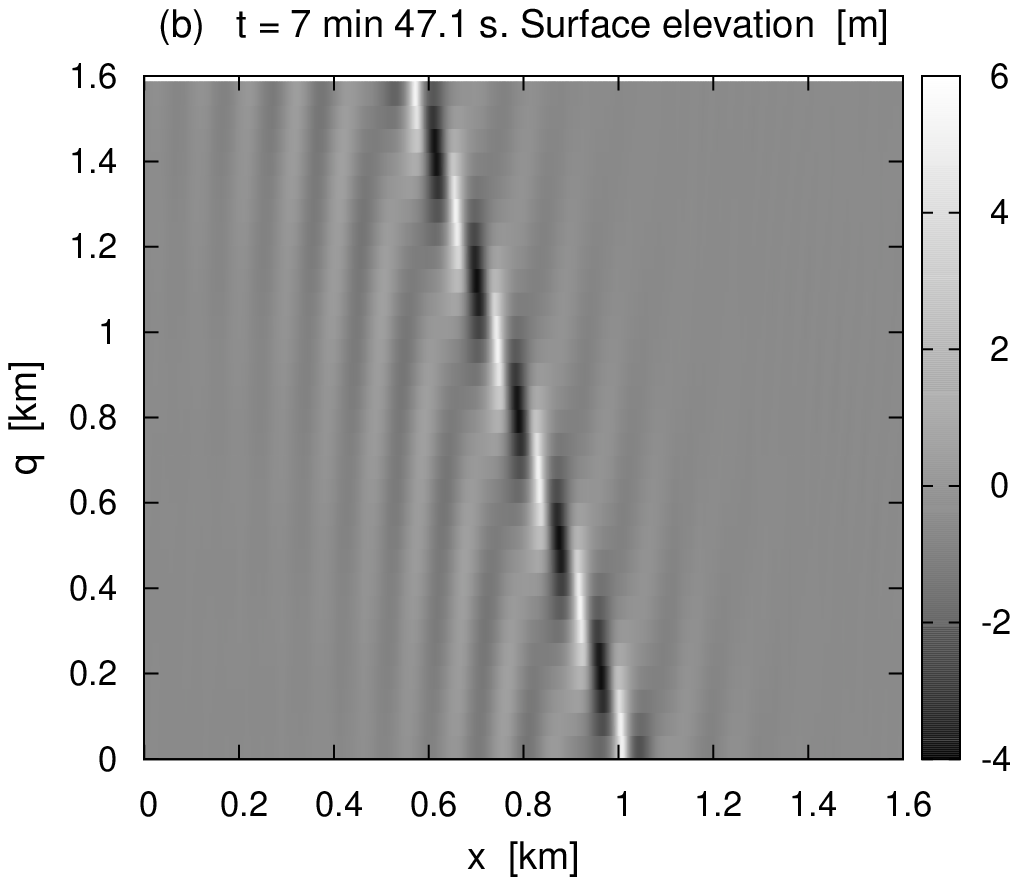,width=91mm}
\end{center}
\caption{Numerical example of formation of rogue waves in a slant wave group.} 
\label{maps} 
\end{figure}

\begin{figure}
\begin{center}
   \epsfig{file=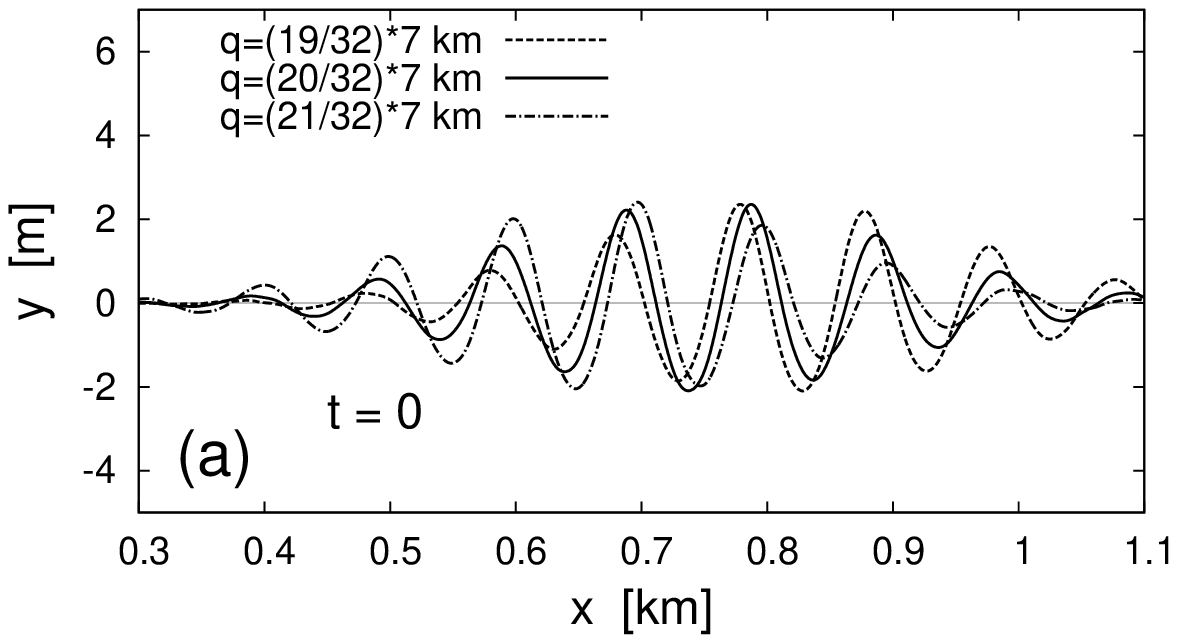,width=86mm}\\
   \epsfig{file=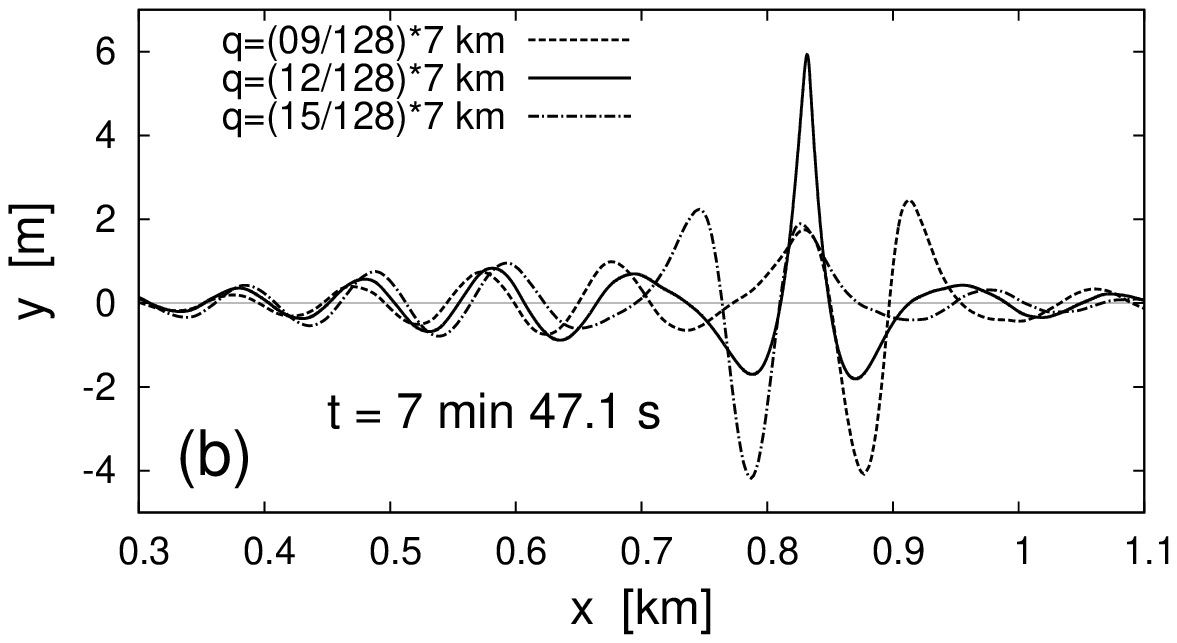,width=86mm}
\end{center}
\caption{Some wave profiles corresponding to Fig.1.
The rogue waves concentrate the energy, and therefore amplitude of the remaining 
waves in the group is decreased.} 
\label{profiles} 
\end{figure}

For convenience of graphical presentation we choose $\lambda=100$ m, 
so the corresponding wave period is $T=\sqrt{2\pi\lambda/g}\approx 8$ s.
The computational domain has the rectangular shape $L_x\times L_q$, with the periodic 
boundary conditions in both directions, and $L_x=2$ km.
For the parameter $L_q$, several different values were taken ($L_q = 4, 5, 6, 7, 8$ km) 
in order to ensure the quasi-2D regime for different angles $\theta$,
at least in an initial stage of evolution; by the way, $\gamma=\arctan(L_x/L_q)$. 
Exception is for small $\theta$, when $L_q=L_x$, and $\gamma=0$.
For example,  in Fig.1 shown are two sub-regions 
of the whole domain 2 km $\times$ 7 km.

The simulations were performed on modern personal computers using numerical method 
described in \cite{RD2005PRE}. The final resolution was about $12000\times3000$ points
in the cases when extreme waves evolved closely to breaking (the beginning of the 
breaking is characterized by a rapid increase of the maximal wave steepness after reaching 
a critical value $s_*\approx 0.5$ rad, which is about the steepness of the limiting 
Stokes wave). Some of the obtained numerical results are presented in the figures. 
In particular, Fig.1a shows a map of the free surface at $t=0$, 
for $s=0.14$, $N\approx 6$, and $\theta\approx 18.4^o$,
while Fig.1b is a map for a later time moment, about several tens of wave periods,
when rogue waves and deep troughs form a specific slant structure resembling 
wake waves after a ship. It is worth noting that fragments of similar wave stripes 
develop spontaneously in nonlinear stage of the modulational instability, 
where they form zigzag patterns, with rogue waves arising mainly at zigzag turns
\cite{R2007PRL,R2010EPJST}. In Fig.2 presented are some wave profiles from Fig.1, which
emphasize amazing features of rogue waves, such as their strong localization 
and a high relative amplification. In general, a highly nonlinear nature of the rogue
wave phenomenon is confirmed in our numerical experiments.

\begin{figure}
\begin{center}
   \epsfig{file=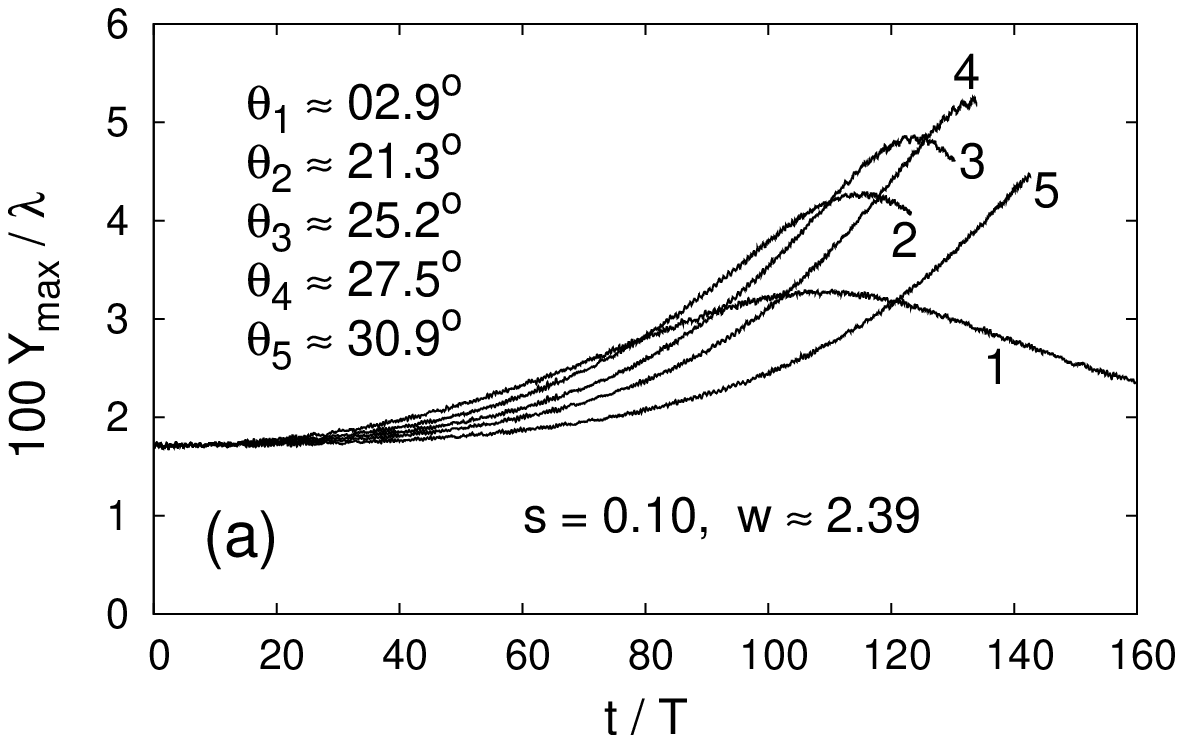,width=82mm}\\
   \epsfig{file=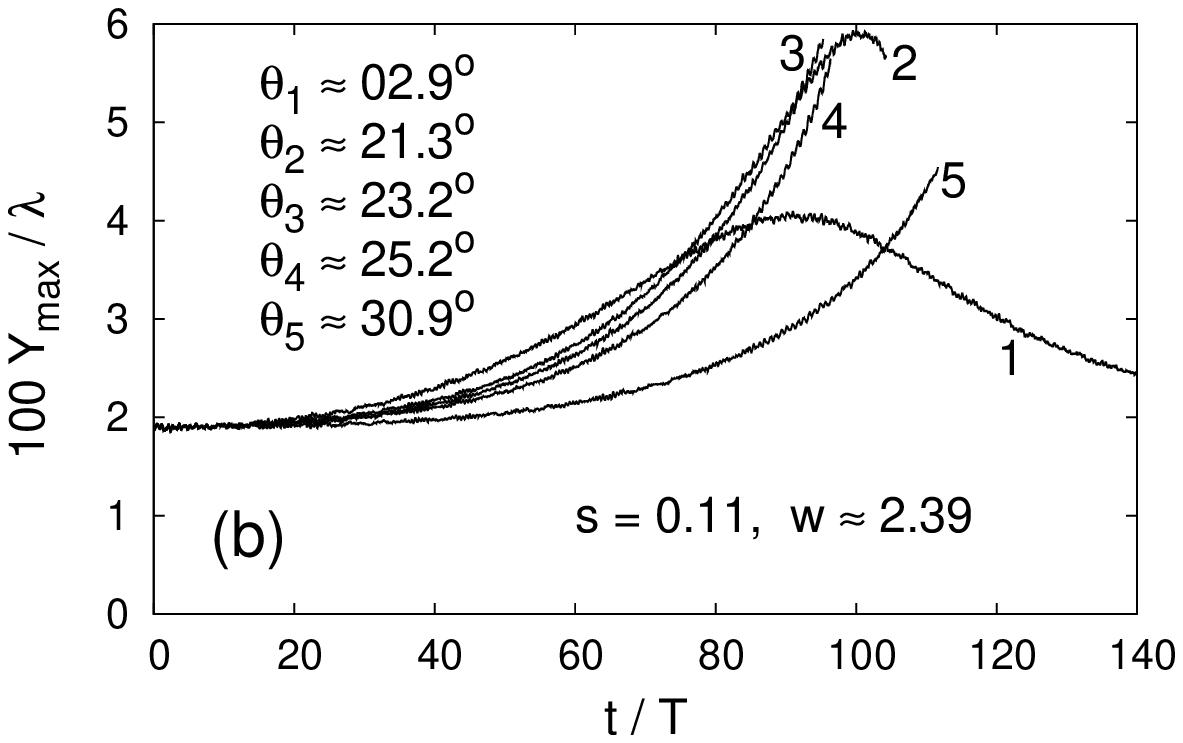,width=82mm}\\
   \epsfig{file=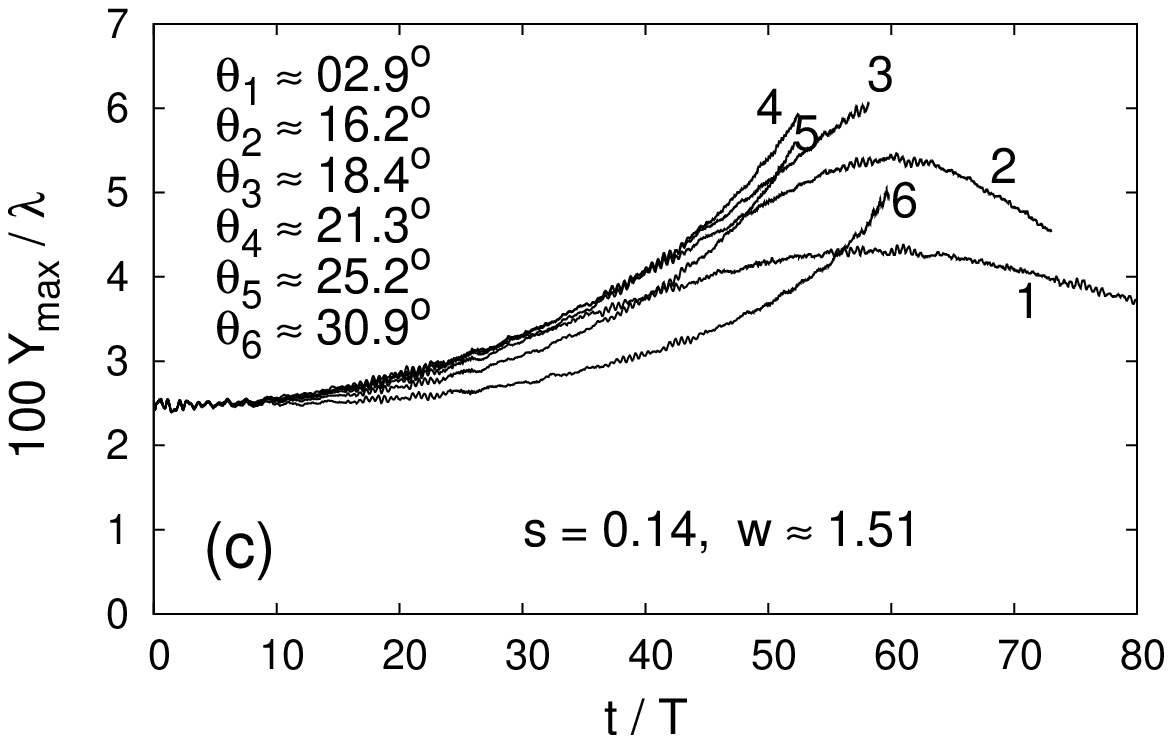,width=82mm}\\
   \epsfig{file=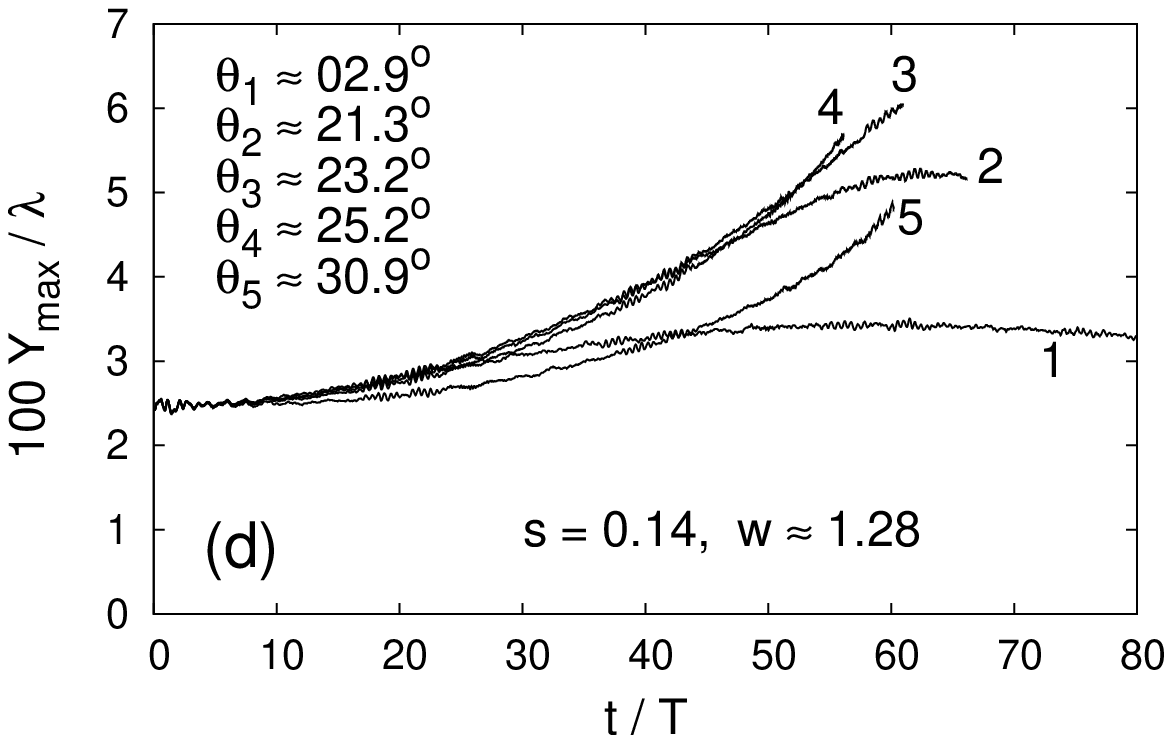,width=82mm}
\end{center}
\caption{Maximum elevation of the free surface versus time in slant wave groups 
with fixed width and steepness, for different angles $\theta$.  
Waves shown in Fig.1b and Fig.2b correspond to dependence `3' in Fig.3c,
slightly before the end point. Wave breaking takes place at the end of experiments  
a5, b3, b4, b5, c3, c4, c5, c6, d3, d4, and d5.} 
\label{Ymax} 
\end{figure}

Very interesting and important is Fig.3, where dependences of the maximum surface 
elevation versus time for different $\theta$ are compared. It is clearly seen that 
for $s$ and $w$ satisfying the relation $4sw\approx 1$ (nearly critical BFI), 
the most ``perfect'' rogue waves with $Y_{\rm max}\approx 0.06\lambda$ arise if 
$\theta$ is sufficiently large, about $18\cdots 28^o$, while for small $\theta$ just 
a moderate wave growth takes place, with subsequent decrease. It should be noted here
that in the absence of wave breaking, after the decrease, the second rise is observed 
at later times (not shown), but in this work we do not discuss that quasi-recurrent 
behavior caused by complex interaction of quasi-solitons constituting the wave group.

An optimal angle $\tilde\theta(s,w)$ increases as $s$ or $w$ decreases 
(compare Fig.3a to 3b, and 3c to 3d).
However, there exists a critical value $\theta_*=\arctan(1/\sqrt{2})\approx 35.3^o$, 
at which the second-order dispersive coefficient of the corresponding 1D NLSE,
approximately describing the dynamics of the wave envelope $B(x_1,t)$ in a moving frame
of reference, changes the sign (see \cite{Zakharov68,R2009PRE} for details):
 \begin{equation}\label{NLSE_1D_theta}
\frac{i}{\omega}\frac{\partial B}{\partial t}=
\frac{[\cos^2\theta-2\sin^2\theta]}{8k^2}
\frac{\partial^2 B}{\partial x_1^2}+\frac{k^2}{2}|B|^2B.
\end{equation}
Our simulations show that actually for $\theta$ approaching $\theta_*$, 
the most tall waves in the stripe become significantly shorter than $\lambda$, 
and finally they break with $Y_{\rm max}$ well below the value $0.06\lambda$.
It should be also noted that for $\theta$  close to $\theta_*$ the quasi-2D regime 
is violated after a short time (crest orientation is then strongly varied across the 
stripe; not shown), and therefore that parametric region cannot be 
accurately investigated with the help of the quasi-2D model.

Qualitatively, such enhanced growth of extreme waves in slant wave groups can be 
explained by noting that with $\theta\not =0$ one should modify the condition (\ref{BFI})
for rogue wave formation as follows,
\begin{equation}
sN/\sqrt{\cos^2\theta-2\sin^2\theta}\gtrsim I_c,
\end{equation}
since the spatial scale perpendicular to the stripe is formally renormalized
in the 1D NLSE by the above square root, as it is clear from Eq.(\ref{NLSE_1D_theta}). 
However, Eq.(\ref{NLSE_1D_theta}) is not applicable with small values of the root, 
because the necessary condition of spectrally narrow wave field 
is then violated very soon in the course of evolution. Adding higher-order linear
dispersive terms into Eq.(\ref{NLSE_1D_theta}), one cannot improve situation, 
since the rapid widening 
of wave spectrum  is a real physical effect for $\theta$ close to $\theta_*$, 
and thus the nonlinear term in NLSE should as well be modified to a non-local 
form determined by the 4-wave matrix element of Zakharov equation \cite{Zakharov68,Z1999}. 
We do not write here the corresponding 1D reduction of the Zakharov equation for slant 
wave stripes, since it is too difficult for analytical treatment, and besides that, 
it is not accurate for large wave amplitudes. So, in the absence of reliable 
analytical estimates for dependence of the maximum surface elevation on $s$, $N$, 
and $\theta$, the above presented numerical results are quite valuable.

To summarize, it has been shown in this work for the first time that 
an oblique orientation of wave fronts in oblong water-wave groups is able to
enhance the process of formation of rogue waves. This 3D effect is most prominent 
when the Benjamin-Feir Index of the wave field is slightly below its critical value.

These investigations were supported by the Russian Foundation for
Basic Research (project no. 09-01-00631),
by the Council of the President of the Russian Federation for Support
of Young Scientists and Leading Scientific Schools (project no. NSh-6885.2010.2),
and by the Presidium of the Russian Academy of Sciences (program 
``Fundamental Problems of Nonlinear Dynamics'').

\end{document}